\documentclass[12pt,aps,pre,amsmath,amssymb,superscriptaddress,floatfix]{revtex4}

\usepackage{graphicx}
\usepackage{bm}
\usepackage[utf8]{inputenc}
\usepackage[T1]{fontenc}

\newcommand{\be}{\begin{equation}}
\newcommand{\ee}{\end{equation}}
\newcommand{\beqn}{\begin{eqnarray}}
\newcommand{\eeqn}{\end{eqnarray}}

\begin{document}

\title{Emergence of disconnected clusters in heterogeneous complex systems}

\author{Istv\'an A. Kov\'acs}
\email{istvan.kovacs@northwestern.edu}
\affiliation{Northwestern University, Department of Physics and Astronomy, Evanston, 60208, USA}
\affiliation{Wigner Research Centre for Physics, Institute for Solid State Physics and Optics, Budapest, 1121, Hungary}

\author{R\'obert Juh\'asz}
\affiliation{Wigner Research Centre for Physics, Institute for Solid State Physics and Optics, Budapest, 1121, Hungary}

\date{\today}

\begin{abstract}
Percolation theory dictates an intuitive picture depicting correlated regions in complex systems as densely connected clusters. While this picture might be adequate at small scales and apart from criticality, we show that highly correlated sites in complex systems can be inherently disconnected. This finding indicates a counter-intuitive organization of dynamical correlations, where functional similarity decouples from physical connectivity. We illustrate the phenomena on the example of the Disordered Contact Process (DCP) of infection spreading in heterogeneous systems. We apply numerical simulations and an asymptotically exact renormalization group technique (SDRG) in $1,2$ and $3$ dimensional systems as well as in two-dimensional lattices with long-ranged interactions. We conclude that the critical dynamics is well captured by mostly one, highly correlated, but spatially disconnected cluster. Our findings indicate that at criticality the relevant, simultaneously infected sites typically do not directly interact with each other. Due to the similarity of the SDRG equations, our results hold also for the critical behavior of the disordered quantum Ising model, leading to quantum correlated, yet spatially disconnected, magnetic domains. 
\end{abstract}

\flushbottom
\maketitle

\thispagestyle{empty}

\section*{Introduction}
Correlated clusters emerge in a broad range of systems, ranging from magnetic models to out-of-equilibrium systems \cite{FK, CI, MD}. Apart from the critical point, the correlation length is finite, limiting the spatial separation between highly correlated sites, leading to spatially localized, finite clusters. Although at the critical point the correlation length diverges, our traditional intuition  is driven by percolation processes, indicating that highly correlated critical clusters remain connected, while broadly varying in their size. As an alternative scenario, we show that at the critical point, highly correlated clusters can break up into spatially disconnected regions in both classical and quantum systems. As a proof-of-concept example, we focus on the Disordered Contact Process (DCP), a simple (out-of-equilibrium) infection spreading model \cite{noest, noest2, moreira}. The critical behavior of the model is well understood in the presence of disorder, at least at the level of statistical properties averaged over a large numbers of samples \cite{moreira,vd,vfm,vojta_3d,vojta_rev}. The detailed simulation of individual samples is much more challenging, due to large dynamical fluctuations and extremely slow dynamics around criticality \cite{hiv, hiv2}.

As an alternative, efficient approach, the strong disorder renormalization group (SDRG) method provides asymptotically exact results for the critical contact process in the presence of disorder, at least below $d=4$ spatial dimensions \cite{hiv,hiv2,im, im2}. Besides being computationally efficient, the SDRG method also offers some counter-intuitive insights into the underlying correlation structure. Namely, the SDRG predicts that highly correlated sites at long time scales (simultaneously infected individuals) typically do not know each other directly, only via indirect connections through the rest of the contact network. According to the SDRG, only a few of these highly correlated, but essentially disconnected, clusters govern the large-scale behavior of the system. Observing these clusters in simulations is notoriously difficult due to extremely slow dynamics, inducing anomalously large fluctuations and poor statistical properties. As a key step, here we show how to find the highly correlated sites efficiently via a quasi-stationary simulation. The simulated density profile is then found to be in a good agreement with the SDRG predictions, confirming that the asymptotic dynamics is governed by spatially disconnected clusters in stark contrast to traditional intuition.

\subsection*{The disordered contact process}

\begin{figure}[h]
\begin{center}
\includegraphics[width=5in,angle=0]{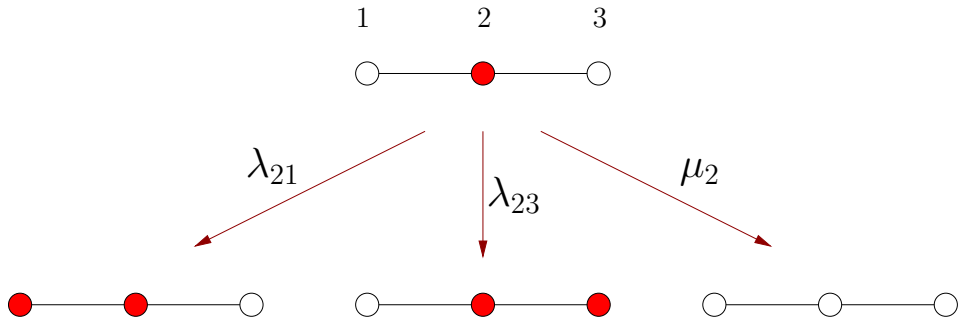}
\end{center}
\vskip -.5cm
\caption{
\label{cp_fig} 
\textbf{Illustration of allowed transitions in the contact process.} 
Allowed transitions of an active (infected) site and two of its neighbours, connected by two links ($12$ and $23$). Active (inactive) sites are depicted by red (white) dots. The corresponding transition rates are written on the arrows. 
}
\end{figure}

The contact process, in the most general case, is defined on a network given by an adjacency matrix $A_{ij}$. As a special case, the network is often chosen to be a d-dimensional hypercubic lattice with nearest-neighbour edges. The state of the system is given by a set of binary variables, $n_i=0,1$, characterizing the sites of the network.    
As the contact process is frequently interpreted as a simple epidemic spreading model, sites with $n_i=1$ are referred as 'infected', while sites with $n_i=0$ are 'healthy' or 'susceptible'. 
The contact process is a continuous-time Markov process on this state space, specified by the rates of possible (independent) transitions, which are the following (for an illustration see Fig. \ref{cp_fig}).

First, infected sites become spontaneously healthy with a rate $\mu_i$, which may be site-dependent.  
Second, infected sites attempt to infect their adjacent sites with rates $\lambda_{ij}$, and the trial is successful if the target site $j$ was healthy. 
Again, the infection rates $\lambda_{ij}$ can vary from link to link. 
We will assume that the infection rate from site $i$ to site $j$ are the same as that from site $j$ to site $i$, i.e. $\lambda_{ij}=\lambda_{ji}$. This variant of the model is also known as the susceptible-infected-susceptible (SIS) model \cite{barrat}. 
This technical restriction, which is necessary for the applicability of the SDRG method, is irrelevant from the point of view of universal critical properties \cite{juhasz2013}. 

The contact process exhibits a continuous, non-equilibrium phase transition between a non-fluctuating (absorbing) phase in which all sites are healthy and a fluctuating phase, where the density of infected sites is non-zero \cite{liggett}. For regular lattices and uniform transition rates the transition falls into the robust universality class of directed percolation \cite{md,odor,odor2,hhl}.  

If the transition rates are independent, random variables, the universality class is well characterized in one, two- and three-dimensional regular lattices due to large-scale Monte Carlo simulations \cite{moreira,vd,vfm,vojta_3d,vojta_rev}. The observed critical behavior is in line with the results of the strong-disorder renormalization group method \cite{hiv,hiv2,im,im2}. According to this, the dynamical scaling of average quantities is ultra-slow, characterized by power-laws of the logarithm of time rather than the time itself, whereas the static critical exponents are also altered compared to the directed percolation class \cite{hiv,hiv2,2d}. Most interestingly, the critical exponents are independent of the form of the distribution of transition rates.    
Besides low dimensional regular lattices, a similar behavior has been found in spatially embedded networks with long-range connections \cite{munoz,munoz2,jk}. These networks consist of a regular, $d$-dimensional, hypercubic lattice and a set of long links, which exist with a probability $p_{ij}\sim \beta d(i,j)^{-s}$, where $d(i,j)$ is the Euclidean distance between $i$ and $j$. 
Note, that even for uniform transition rates these networks contain a so-called  topological disorder due to the random connectivity of sites by long links.
In the case $s=2d$, the topological dimension is finite and varies with $\beta$ \cite{bb,coppersmith,juhasz2012}. As it was demonstrated in Refs. \cite{munoz,munoz2,jk} for $d=1$ and $s=2$, the contact process shows an ultra-slow scaling and the critical exponents vary with $\beta$, while an additional disorder in the transition rates is irrelevant. According to a general scaling argument presented in Ref. \cite{jki}, a qualitatively similar behavior is expected in higher dimensions for $s=2d$.    

\subsection*{Strong-disorder renormalization group}

The strong-disorder renormalization group is the key method to understand the long-time behavior of the disordered contact process (DCP). For a general review and a detailed introduction we refer the reader to Ref. \cite{im,im2}. The first application of the method to the DCP was in Ref. \cite{hiv,hiv2}. Next, we recapitulate the essential features of the method. 

\begin{figure*}[ht]
\begin{center}
\includegraphics[width=6in,angle=0]{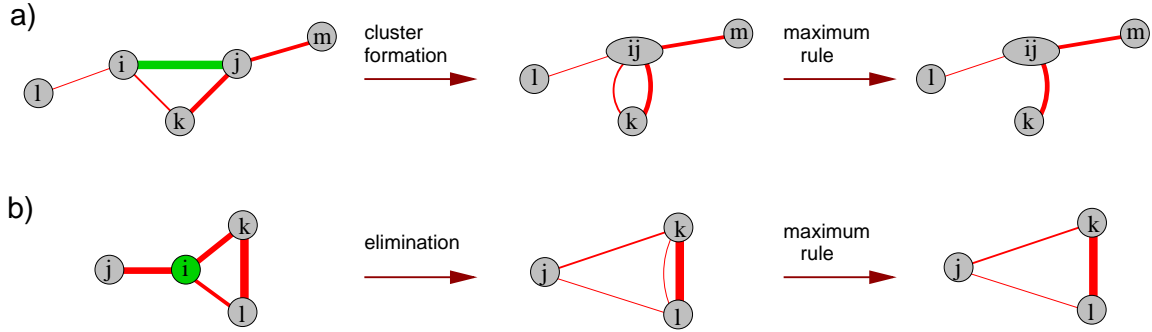}
\end{center}
\vskip -.5cm
\caption{
\label{sdrg_fig} 
\textbf{Illustration of the SDRG method.}
Two types of reduction steps of the SDRG procedure: formation of a cluster (a) and elimination of a site (b). The thickness of lines indicates the magnitude of the infection rate on the corresponding link. The link (a) and the site (b) to be decimated is shaded in green. For details, see the text. 
}
\end{figure*}

The SDRG is a sequence of iterative steps operating on blocks of sites containing the largest rate (either an infection rate or a healing rate) of the model, see Fig. \ref{sdrg_fig}. 
If the largest rate is an infection rate, $\lambda_{ij}$, the block of sites $i$ and $j$ is merged to a cluster, characterized by an effective healing rate: 
\be 
\ln\tilde\mu_{ij} = \ln\mu_i + \ln\mu_j - \ln\lambda_{ij} + \ln 2. 
\label{mueff}
\ee
This simplification is a good approximation if $\mu_i$ and $\mu_j$ are much smaller than $\lambda_{ij}$. 
If the largest parameter is a healing rate, $\mu_i$, site $i$ is deleted, and effective infection rates are generated between all neighbors of site $i$: 
\be 
\ln\tilde\lambda_{jk} = \ln\lambda_{ji} + \ln\lambda_{ik} - \ln\mu_i. 
\label{lambdaeff}
\ee
This approximation is justified under the condition $\lambda_{ji},\lambda_{ik}\ll \mu_i$. 
Apart from a one-dimensional lattice, it may happen that a newly formed cluster has, through its constituents, two infection rates to a third site. Similarly, if a new infection rate is generated, there may be a pre-existing infection rate between those two sites. These situations are usually treated by discarding the smaller infection rate, which is known as the maximum rule. 

The above steps are applied sequentially, lowering thereby gradually the number of degrees of freedom, as well as the rate scale $\Omega=\max\{\mu_i,\lambda_{ij}\}$. Regarding the set of clusters present in the system, the SDRG procedure can also be viewed as a special coagulation-annihilation process with extremal dynamics.  
The critical behavior of the DCP is governed by the so called infinite-randomness fixed-point (IRFP) of the SDRG transformation \cite{hiv,hiv2, fisher,fisher2}, at least for sufficiently strong initial disorder \cite{hiv,hiv2, vd, nft,nft2, hoyos}. As the IRFP is approached along the critical renormalization trajectory, the distribution of logarithmic rates broadens without limits, and both types of reduction steps become asymptotically exact. Furthermore, this feature also justifies the applicability of the maximum rule, and even the neglection of the $\ln{2}$ term in Eq. (\ref{mueff}).  
In the numerical SDRG calculations we therefore resorted to these approximations, which, as a further advantage, enable a computationally very efficient implementation of the SDRG procedure \cite{3d,3d2}.

Following the SDRG transformation of the DCP at a given realization of random rates down to a rate scale $\Omega\ll 1$ provides an effective DCP with a smaller set of degrees of freedom, which approximates well the dynamics of the original system for times $t\gg \Omega^{-1}$.  
Consider, for instance, the state of the system, evolved from a fully occupied initial state, at some late time $t$. Applying the SDRG procedure down to rate scale $\Omega=t^{-1}$, provides the set of sites with an $O(1)$ occupation as the constituents of the clusters of the renormalized system, while the sites eliminated during the procedure will have an almost zero occupation probability.   

In the case of a finite sample, the true steady state is always the absorbing, empty lattice state. Accordingly, all clusters are removed during the SDRG procedure, the last one at some rate $\Omega_1$, the second last at $\Omega_2>\Omega_1$, in general, the $n$th cluster becomes irrelevant at $\Omega_n>\Omega_{n-1}$. 
The inverse rates, $\Omega_1^{-1},\Omega_2^{-1},\dots$, are interpreted as the mean lifetimes of the corresponding clusters removed during the procedure. 
Clearly, the last one, $\Omega_1^{-1}$, gives the mean lifetime of the sample needed to be absorbed in the empty state. 
In finite but large systems in the vicinity of the critical point, the last few lifetimes such as $\Omega_1^{-1}$ and $\Omega_2^{-1}$ are typically well separated, meaning that they differ by orders of magnitude, and this time-scale separation becomes more and more pronounced with increasing system size. 
Consequently, in typical samples there is a considerable time span, $\Omega_1^{-2}\ll t\ll\Omega_1^{-1}$, within which practically no sites but the constituents of the lastly removed cluster are occupied. 

The structure of the lastly removed cluster can also be used to determine a sample-dependent pseudo-critical point: making a double sized system by glueing together two copies of the original one, the onset of the active phase is indicated when the last cluster is different from that of the original system \cite{2d}.

\section*{Quasi-stationary simulation}

The simulations were implemented for regular lattices in the following way. 
An occupied site ($i$) is randomly chosen and, with a probability $\mu/(\mu+1)$ it is made unoccupied, or, with a probability $1/(\mu+1)$, one of its $n$ neighbors ($j$) is picked with a uniform probability and infected with the probability $\lambda_{ij}$, provided it was healthy.
The time increment related to such a move is $\Delta t=1/N_s$, where $N_s$ is the total number of infected sites. 

For non-regular networks, where the sites may have different coordination numbers, a different implementation is needed in order to correctly simulate the process. Here, besides the list of infected sites, also a list of active links is stored, which contains all directed links with an infected source site. The number of elements of this list is denoted by $N_e$. 
Then, with a probability $\mu N_s/(\mu N_s + N_e)$, a healing event occurs on one of the infected sites, chosen equiprobably, whereas,  
with the complementary probability, $N_e/(\mu N_s + N_e)$, an infection event is attempted. In this case, an active link ($ij$) is chosen with a uniform probability from the corresponding list, and the target site ($j$) is infected with a probability $\lambda_{ij}$, provided it was healthy.  
The time increment of such an elementary move is $\Delta t=1/(\mu N_s + N_e)$. 

We apply a quasi-stationary simulation with a reflecting boundary condition. This means that, at the point where there is only a single active site in the system, the healing event is rejected. This way we only need to validate that the system reached the quasi-stationary state, easily done by checking whether the mean density and its variance remain unchanged under increasing the relaxation time. 
The total simulation time was typically $10^{28}$, the first half of which is left for relaxation, while the measurements are performed in the second half.

\begin{figure*}[ht!]
\begin{center}
\includegraphics[width=6.in,angle=0]{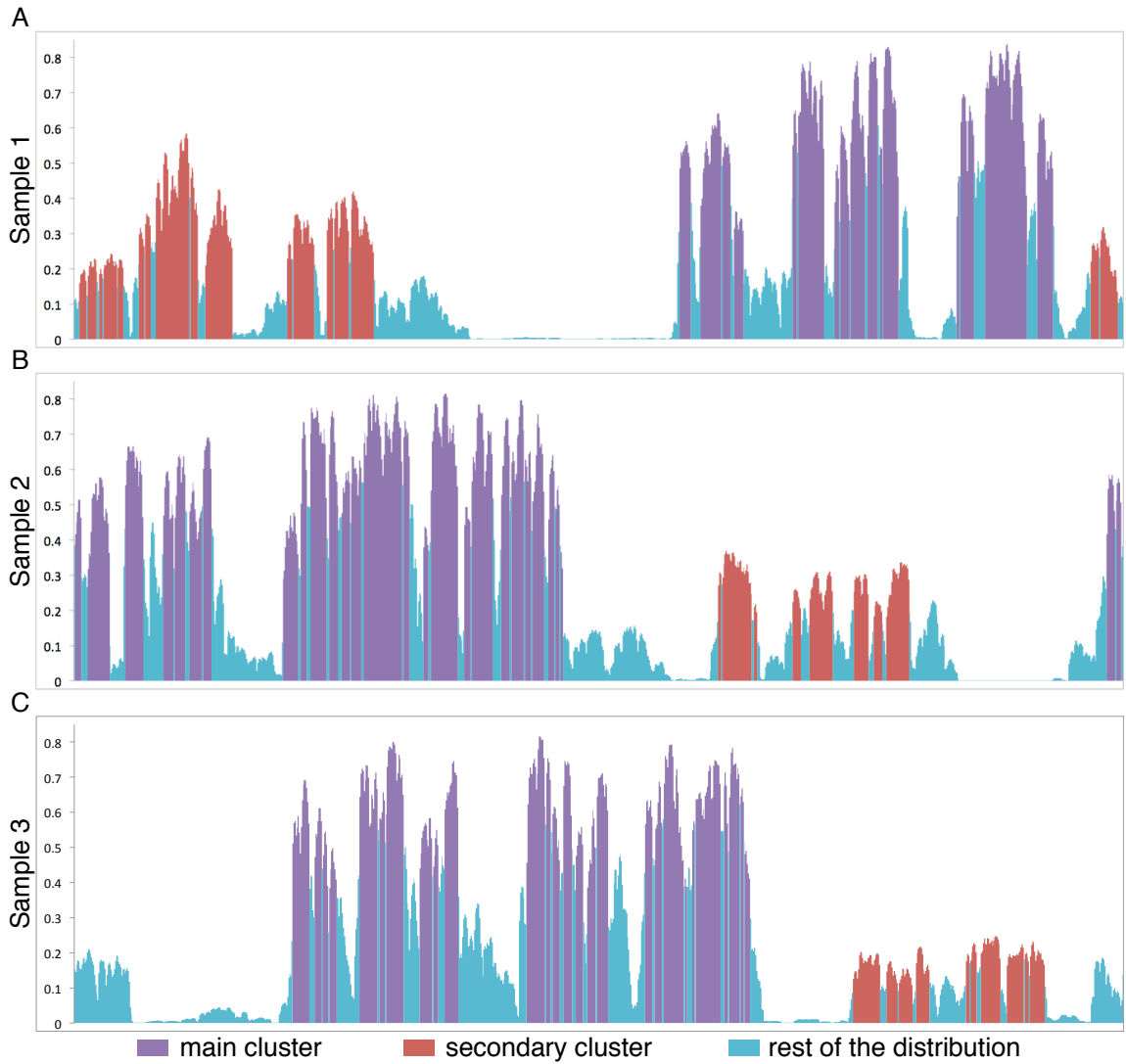}
\end{center}
\vskip -.5cm
\caption{
\label{fig_1d} 
\textbf{Results in 1D systems.} Asymptotic probability of being active in the quasi-stationary simulations as the function of the spatial coordinate for $N=1000$ sites with periodic boundary conditions and strength of disorder $\alpha=1$. As illustrated by the purple and red clusters the SDRG is able to efficiently capture the activity profile even with a few clusters. The correlated clusters are disconnected fractals with a fractal dimension $d_f=\frac{1+\sqrt{5}}{4}\approx 0.819$ \cite{fisher, fisher2}, corresponding to a highly uneven activity profile. Gaps of strongly reduced density inside the clusters can be seen already at small scales, indicating asymptotically disconnected clusters of activity.
}
\end{figure*}

First, in each sample, we determine an individual pseudo-critical point, which, for the system sizes we use, can significantly deviate from the ensemble average.
Following the method proposed in Ref. \cite{ferreira}, we 
identify the pseudo-critical point with the maximum of susceptibility \cite{binder} defined as 
\be 
\chi=N\frac{\langle\rho^2\rangle -\langle\rho\rangle^2}{\langle\rho\rangle},
\ee 
where $N$ is the total number of sites, $\rho$ denotes the global density of active sites, and $\langle\cdot\rangle$ stands for the expected value in the quasi-stationary state.

Having estimated the pseudo-critical point, we performed here quasi-stationary simulations for a longer time, $10^{30}$, and measured the mean local densities (Figs. \ref{fig_1d}-\ref{fig_2dlr}). In addition, we started a number of independent simulations and averaged over them, in order to avoid the (rather improbable) possibility that the long-lasting activity is stuck at a cluster other than the one with the longest lifetime. 

For a satisfying agreement with the SDRG method for moderate system sizes, the strength of the disorder needs to be sufficiently large, i.e. the distribution of the logarithmic infection rates needs to be sufficiently broad. In practice, the infection rates were chosen from a power-law distribution, $P_<(\lambda)=\lambda^{1/\alpha}$, where the exponent $\alpha>0$ controls the strength of disorder.

\section*{Comparison of the cluster structure with the simulations}

\begin{figure*}[ht!]
\begin{center}
\includegraphics[width=7.in,angle=0]{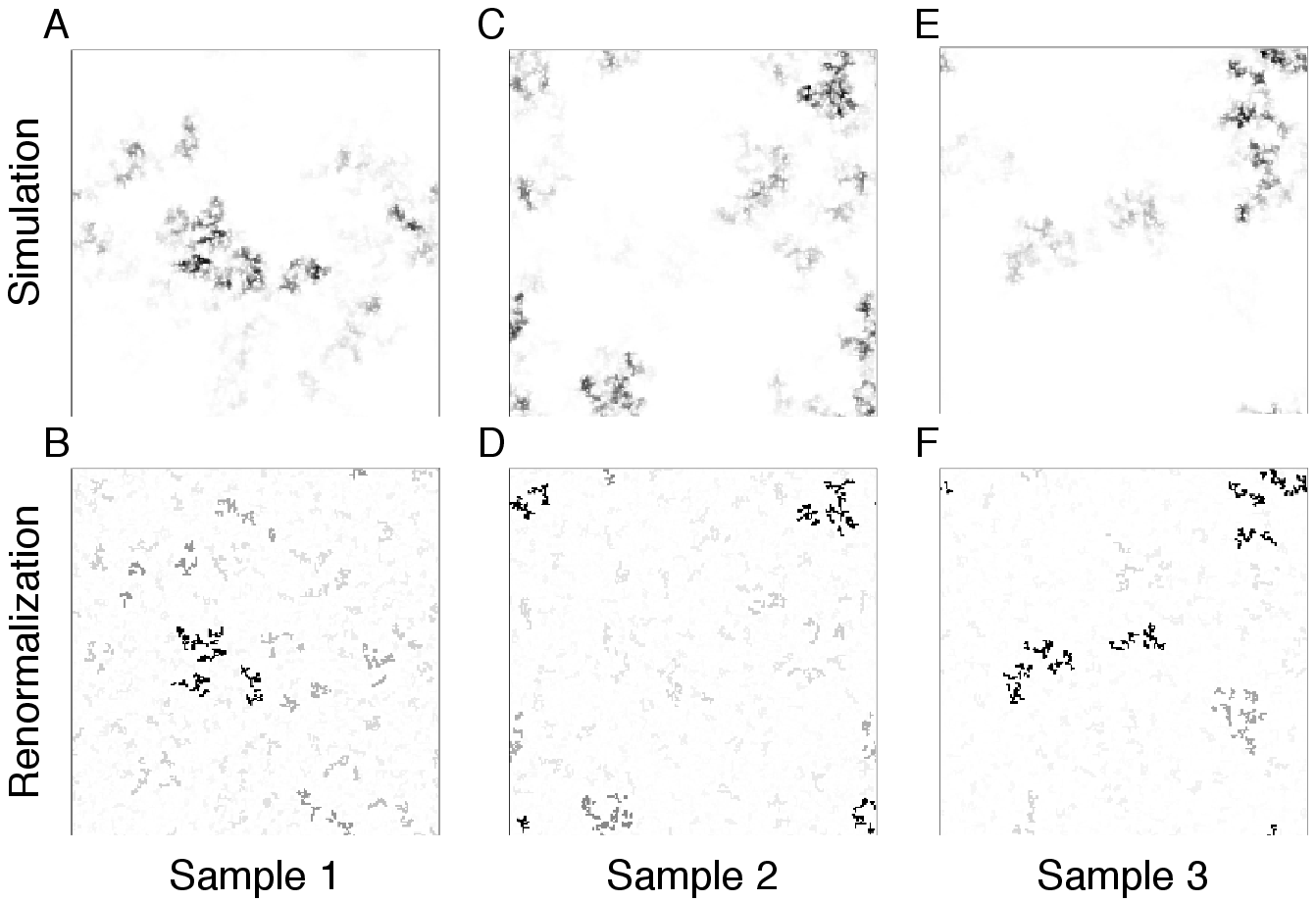}
\end{center}
\vskip -.5cm
\caption{
\label{fig_2d} 
\textbf{Results in 2D systems.} A good agreement is found between the asymptotic density profile in the simulations and the largest clusters obtained by the SDRG method. In 2D the fractal dimension of the clusters is $d_f= 1.02(2)$ \cite{2d}, illustrated here for $N=200\times200$ sites with periodic boundary conditions and strength of disorder $\alpha=4$. The grayscale indicates the site probability of being active in the simulations and the cluster size in the SDRG calculations, respectively. 
}
\end{figure*}

\begin{figure*}[ht!]
\begin{center}
\includegraphics[width=7.in,angle=0]{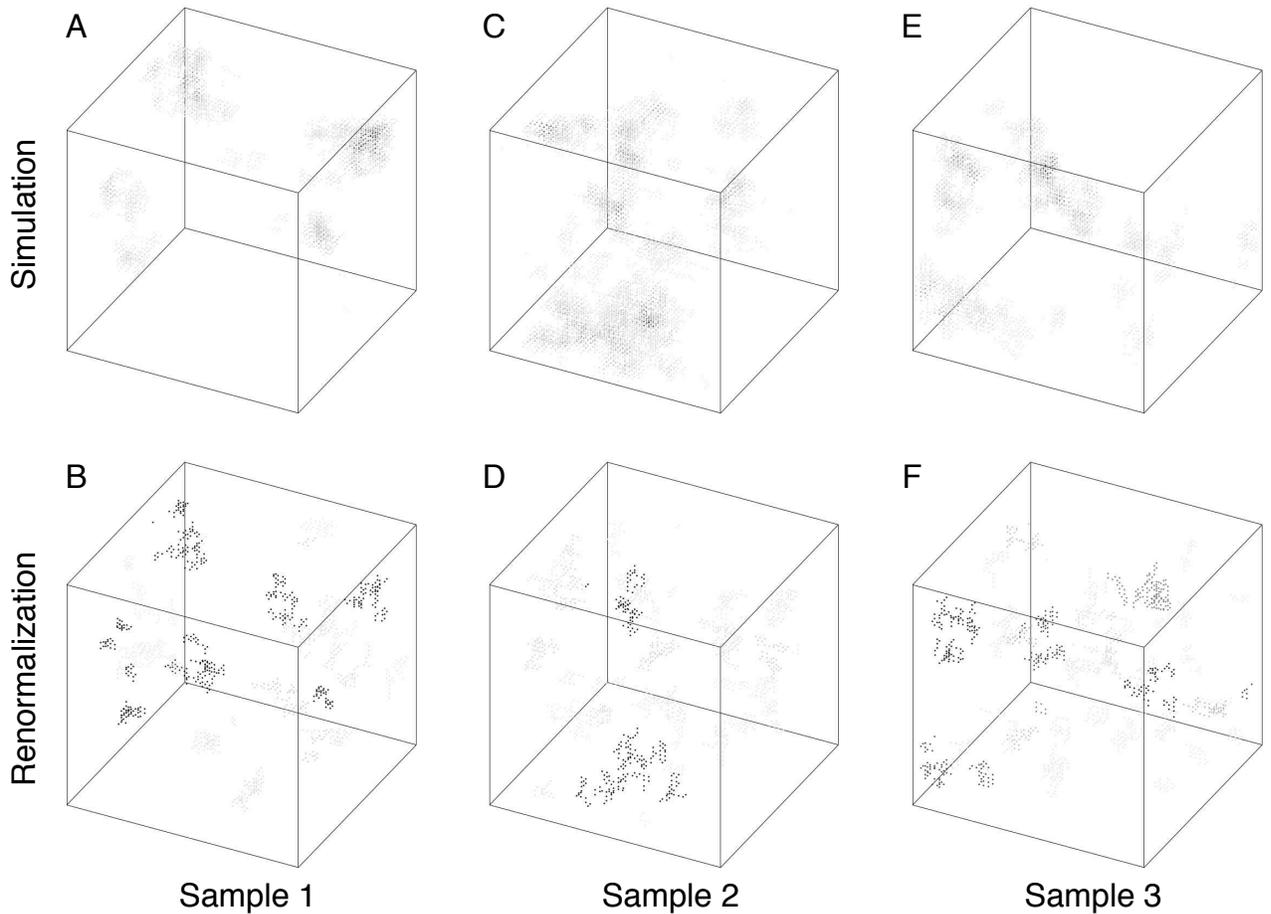}
\end{center}
\vskip -.5cm
\caption{
\label{fig_3d} 
\textbf{Results in 3D systems.} We again find a good agreement between the asymptotic density profile in the simulations and the clusters obtained by the SDRG method. In 3D the fractal dimension is $d_f= 1.16(2)$ \cite{3d,3d2}, illustrated here for $N=50\times50\times50$ sites with periodic boundary conditions and strength of disorder $\alpha=4$. For better visibility, only sites with at least $10\%$ of the maximum probability are indicated in the simulations, while we show only clusters of size above the square root of the size of the main cluster. 
}
\end{figure*}

\begin{figure*}[ht!]
\begin{center}
\includegraphics[width=7.in,angle=0]{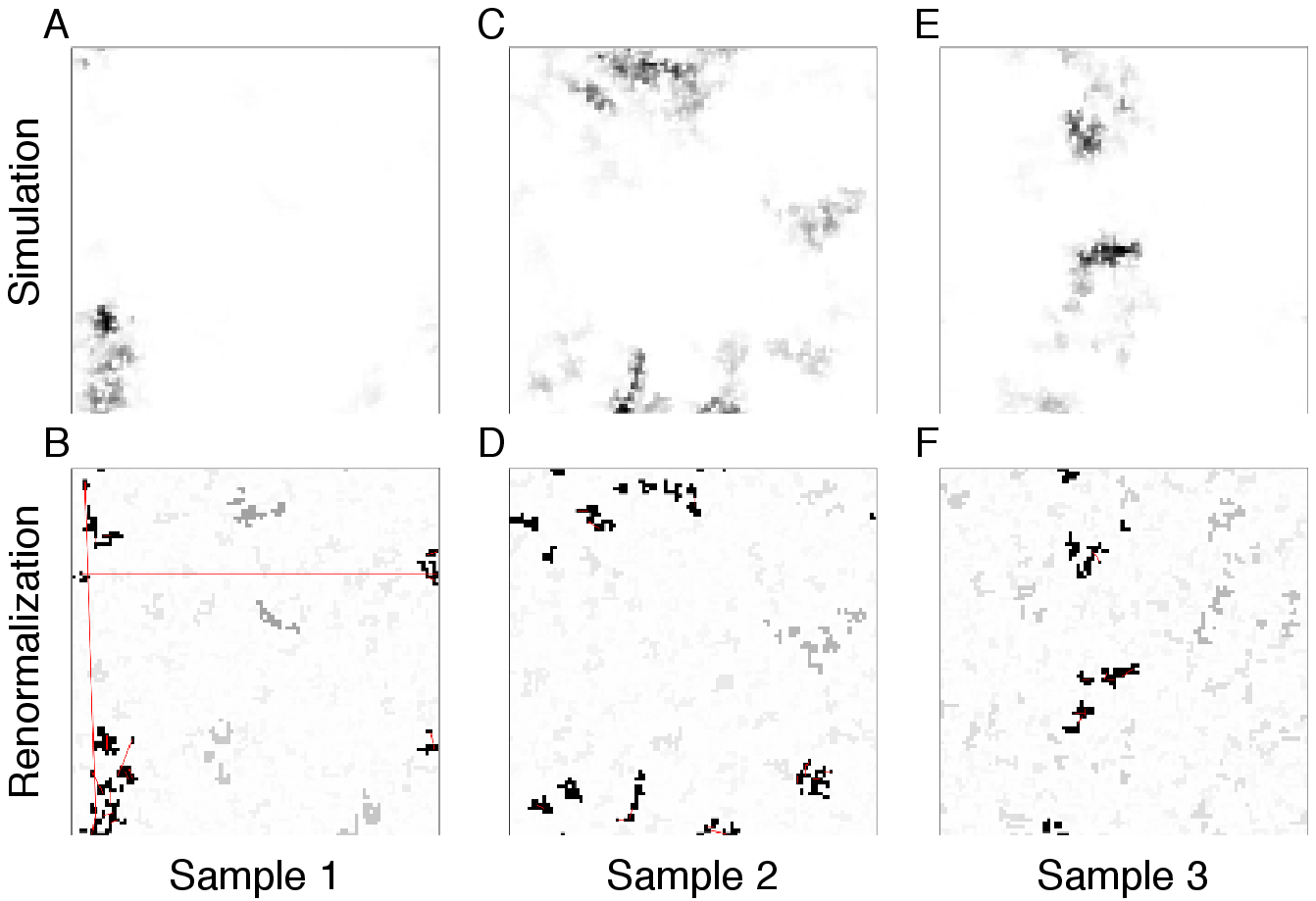}
\end{center}
\vskip -.5cm
\caption{
\label{fig_2dlr} 
\textbf{Results in 2D with long-range interactions.} Here we show our results for algebraically decaying interaction probability with an exponent $s=4$, for $N=100\times100$ sites with periodic boundary conditions. The number of short-cuts is $N/16$ and the strength of disorder is $\alpha=4$. Red links indicate long-range connections within the main cluster. The critical clusters are even more strongly disconnected objects, with a formally zero fractal dimension \cite{jki}.
%
}
\end{figure*}

The average order parameter of the DCP is related to the fraction of original sites comprised by the clusters of the renormalized system, which decreases gradually as the SDRG proceeds. Its critical scaling properties can then be inferred from its dependence on the inverse time scale $\Omega$ and inverse length scale (the number density of clusters) along the critical trajectory of the SDRG \cite{im,im2}. 
Instead of this, here we compare the spatial structure of SDRG clusters with the map of local occupancies obtained by simulations in individual samples.  

Ideally, the comparison of the two methods should be done under the same circumstances: i) for the same random sample (set of random rates), ii) at the same time (given by $\Omega^{-1}$ in the SDRG method). 
The first requirement is, however, not the appropriate choice for a fair comparison of the methods. The reason is that the SDRG transformation, due to its approximative nature at early stages, does not preserve the position of the critical point. For instance, a truly critical initial system will depart from the critical trajectory and, in order to arrive at the critical IRFP, a compensatory initial shift in the control parameter is needed to be imposed. Therefore, rather than the initial point of the renormalization trajectory, its end point has to match the sample used in simulations. For this reason, the first requirement, i.e. the complete identity of rates must be relaxed, but a notion of ``equivalence of the random environment'' must still be preserved. 
Our approach to overcome this controversy was the following. We chose the infection rates randomly, while kept the healing rates constant. 
The latter can then serve as a control parameter and can be used to compensate the shift induced in the SDRG. 
The amount of the shift is quantitatively {\it a priori} unknown, therefore it is natural to choose the critical point as a ``common point'' to which the system can be tuned in both methods by using an indicator of criticality internal to that method. 
By this construction, the transition rates used in the two methods are not completely identical, but the difference is only in the uniform healing rate, while the random infection rates, which form the ``random environment'', and which govern the shaping of SDRG clusters, are identical.

Concerning the second requirement, we implemented the SDRG procedure up to the stage at which only one cluster remains in the system. Provided the lifetimes of clusters are well separated, this corresponds to the time span between the lifetime of the second last and the last cluster. The state of the system in this time span can be conveniently simulated, even without knowing the above lifetimes, by the above described quasi-stationary simulation, which prevents the system from getting absorbed in the empty state.  

In Fig. \ref{fig_1d}, we illustrate for three random samples that the main SDRG cluster (purple) not only contains all sites of nearly maximal probabilities, but also captures the majority of the probability weight. For such, moderately sized, systems smaller clusters still play a role, as shown by probability distribution captured by the secondary cluster. Interestingly, both the main and secondary clusters are fractured by deep gaps in the probability distribution, as signatures of asymptotically disconnected clusters. Overall, we see a good agreement already for moderate system sizes as illustrated in Figs. \ref{fig_1d}-\ref{fig_3d} for $d=1,2$ and $3$ dimensional lattices as well as for a two-dimensional lattice with long-range interactions (Fig. \ref{fig_2dlr}).

\section*{Discussion}

In this paper, we studied the disordered contact process (DCP) in a random environment. By applying the combination of quasi-static simulations and an efficient renormalization group method, we showed that the critical behavior of the DCP is dominated by one strongly correlated cluster. Yet, as opposed to equilibrium systems, the governing clusters are predominantly disconnected objects, indicating that individuals who are infected at the same time typically do not know each other directly. According to the SDRG method, such disconnected clusters emerge from a strong, positive feedback mechanism, in which remote sites can effectively infect each other over and over again for a prolonged amount of time through indirect paths in the contact network \cite{phd}. 

Besides infection spreading, variants of the DCP have recently gained increasing interest also in functional modeling of the brain \cite{moretti}. Our results suggest that, as opposed to traditionally expected functional 'blobs', strongly correlated brain regions are not necessarily directly connected. In other words, functional connectivity might qualitatively deviate from physical connectivity. This expectation exists in addition to the challenge that in the brain correlations do not necessarily decay with increasing physical distance \cite{brain_review}. 
We leave the extension of our results to dynamical models on experimentally obtained brain connectome datasets for future studies.


\section*{Acknowledgements}

This paper was supported by the National Research, Development and Innovation Office - NKFIH under grants No. K128989 and No. K131458. IAK was supported 
by the 'Nation's Young Talents Scholarship', Ministry of National Resources, Hungary under Grant No. NTP-NFT\"O-17-C-0247.

\section*{Author contributions statement}
I.A.K initiated the project and performed the SDRG calculations. R.J. performed the MC simulations. All authors contributed to writing the manuscript.

\section*{Competing interests}
The authors declare no competing interests.

\section*{Data availability}
All data generated or analysed during this study are included in this published article.


\end{document}